# SPATIAL DISCRETE CHOICE AND SPATIAL LIMITED DEPENDENT VARIABLE MODELS: A REVIEW WITH AN EMPHASIS ON THE USE IN REGIONAL HEALTH ECONOMICS

Billé, Anna Gloria[1]

and

Arbia, Giuseppe[2]

**ABSTRACT** Despite spatial econometrics is now considered a consolidated discipline, only in recent years we have experienced an increasing attention to the possibility of applying it to the field of discrete choices (*e.g.* Smirnov, 2010 for a recent review) and limited dependent variable models. In particular, only a small number of papers introduced the above-mentioned models in Health Economics. The main purpose of the present paper is to review the different methodological solutions in spatial discrete choice models as they appeared in several applied fields by placing an emphasis on the health economics applications.

## 1. INTRODUCTION

The birth and the development of the Spatial Econometric field is historically due to Pealink and Klaassen (1979) with the publication of the first volume entitled "Spatial Econometrics", in which a particular emphasis was devoted to funding principles of the discipline. About ten years later, Anselin (1988) defined the newly born field as "the collection of techniques concerning the peculiarities caused by space in the statistical analysis of models on regional sciences". These peculiarities, that is the spatial features of data, do not allow the use of standard econometric techniques. Many researchers have dealt with some peculiar spatial analytic issues like *spatial heterogeneity* (*e.g.* Anselin, 1988, 1990; McMillen, 1992; Fotheringham *et al.*, 1998; McMillen and McDonald, 2004; LeSage, 2004; among many others), *spatial autocorrelation* (*i.e. spatial dependence*) (*e.g.* Anselin, 1988; McMillen, 1992; Anselin and Florax, 1995; Anselin, 2002; Wang *et al.*, 2013; among many others) and *spatial heteroskedasticity* (*e.g.* Case, 1992; McMillen, 1992; Pinkse and Slade, 1998; Pinkse *et al.*, 2006; Bhat and Sener, 2009; among many others), and most of these researchers are still trying to deal with all of these particular spatial issues at the same time, since it is usually the case that the presence of one of more of them implies the presence of the others. In particular, some of them have also highlighted the differences between spatial dependence and time-series dependence (*e.g.* Besag, 1974; Arbia, 2006; Wang *et al.*, 2013; among many others). Recently, several review papers on Spatial Econometrics in different fields were published. Regarding the development and the diffusion of Spatial Econometrics in different disciplines, see for example Anselin (2007) about the Regional Science field, Bell and Dalton (2007) on microeconomic spatial effects, Holloway *et al.* (2007) on bio-economics and land use spatial models, Anselin (2010) about the thirty years of Spatial Econometrics, and a last most recent of Arbia (2011) about the five years of the Spatial Econometrics Association.

[1] PhD candidate, Department of Economics (DEc), University G. d'Annunzio of Chieti-Pescara, 65126 Italy.
*E-mail address*: ag.bille@unich.it.
[2] Full Professor of Statistics, Department of Statistics, Catholic University of the Sacred Heart, Rome, 00198 Italy.
*E-mail address*: giuseppe.arbia@rm.unicatt.it.

Despite Spatial Econometrics is now considered a consolidated discipline, only in recent years we have experienced an increasing attention to the possibility of applying it to the field of Discrete Choices (*e.g.* Smirnov, 2010 for a recent review) and Limited Dependent Variable Models. For the last one, an interesting empirical example is that of Mizobuchi (2005). As a matter of fact, although the consideration of spatial interaction models is becoming prevalent in many empirical applications with a continuous dependent variable, these aspects are still rarely accounted for in Discrete Choice and Limited Dependent Variable Models. In particular, only a small number of papers introduced the above-mentioned models in Health Economics. In most cases, the additional information coming from the geographical location of data is relevant in order to inform politicians on particular socio-economic phenomena in an accurate way. For example, it is of political interest to describe the entity of regional inequalities on the total number of general practitioners available for each region, since the more this entity is significant the more are these regional inequalities in the access on Health Services (*see e.g.* Bolduc *et al.*, 1996). Moreover, spatial analysis can lead to the association of two or more phenomena under study. For example, locations with a high concentration of individuals affected by a particular respiratory disease can be correlated with a high level of atmospheric pollution (*see e.g.* Best *et al.*, 2000). The individuation of these high-pollution locations can help political choices to focused and localized objectives of prevention and reduction of the atmospheric pollution. As a consequence, the first main purpose of this paper is to review the different methodological solutions in Spatial Discrete Choice Models as they appeared in several applied fields by placing an emphasis on Health Economics applications.

A plausible reason for the relatively scarce diffusion of these models is certainly their complexity (*e.g.* Holloway *et al.*, 2002; Mizobuchi, 2005), often requiring a multidimensional integration to estimate the parameters with Full ML approach (the so-called *multidimensional integration problem*, Fleming, 2004). Pinkse and Slade (1998) proposed a GMM approach for spatial error probit models, and Klier and McMillen (2008) proposed a Linearized version of Pinkse and Slade's GMM to avoid the GMM's infeasible problem of inverting $n$ by $n$ matrices. When we specify a spatial autoregressive probit model we generally derive the maximum likelihood function implied by the *reduced form* of the spatial model. A major problem in maximizing the log-likelihood function of the spatial model is represented by the fact that it repeatedly involves the calculation of the determinant of an n by n matrices (which are related to the weighting matrix) whose dimension depends on the *sample size*. On the other hand, the lack of consideration of spatial relationships, especially in cross-sectional studies, usually causes distortion and/or inconsistency on the usual estimators (*e.g.* Case, 1992). Furthermore, from a substantive point of view, spatial parameters usually bear an important information content, so that they cannot be thought of as simply *nuisance parameters*. In fact, spatial dependence not only means lack of independence between observations, but also a spatial structure underlying these spatial correlations (Anselin and Florax, 1995). Traditionally, spatial models with continuous dependent variables are estimated by maximum likelihood (ML) method (Anselin, 1988; Arbia, 2006). However, as we said before, the ML approach can be *computationally very heavy* when dealing with spatial models with discrete or limited dependent variables. Although a problem of computational burdensomeness exists, the primary advantage of the ML estimator is the potential for efficiency (Wang *et al.*, 2013). Therefore, in a related work (Arbia and Billé, forthcoming) we have also simulated a spatial mixed autoregressive-regressive probit model in order to compare GMM estimation and Linearized GMM estimation with ML estimation in terms of their computational times and their statistical properties.

During recent years, some methodological and computational solutions have been proposed (e.g. Fleming, 2004; Pinkse et al., 2006; Klier and McMillen, 2008; Bhat and Sener, 2009; Wang *et al.*, 2013) and, due to the possible computational advantages, many researchers are increasingly incline to use Bayesian inference and its computational algorithms (*e.g.* MCMC, Gibbs sampling) in Spatial Discrete Choice Models (*e.g.* LeSage, 2000; Holloway *et al.*, 2002; Rathbun and Fei, 2006; among many others). On the other hand, a series of papers with semiparametric or nonparametric approaches are increasing in Spatial Econometrics (*e.g.* McMillen and McDonald, 2004).

Despite the relatively importance of Spatial Panel Econometrics especially in the last years, we exclude from the present work the analysis of the papers with panel datasets, with the exception of some examples, and we concentrate on the analysis of cross-sectional datasets. Therefore, we focus our research on Spatial Discrete Choice and Limited Dependent Variable Models, in particular on spatial binary models and their estimation problems, with applications on Health Economics and the use of cross-sectional data.

The paper is structured in the following way. Section 2 presents an overview of spatial discrete choice models mainly based on classical inference with no concern about aspatial dependence. In particular, Section 2.1 will be focused on models with a dichotomous dependent variable (*i.e. spatial binary probit/logit models*) with a discussion on the statistical and computational properties of the different estimators, Section 2.2 on models with a dependent variable that takes more than two outcomes ordered between them (*i.e. ordered probit/logit models* and their variants), Section 2.3 on models with a dependent variable that takes more than two outcomes unordered between them (*i.e. multinomial probit/logit models* and their variants), and, finally, Section 2.4 on models for count data (*i.e. spatial Poisson/negative binomial models* and their variants). Section 3 discusses the same spatial discrete choice models but following a Bayesian approach (*i.e. Bayesian spatial discrete choice models*). In some of the above sections occasionally we will also refer to papers on discrete choice with no spatial effects, but with an emphasis on regional health economics applications. In this way we will be able to highlight the specific problems emerging in the area and to suggest possible spatial econometric solutions. In Section 4 we introduce different limited dependent variable models. Section 5 introduces some of the spatial discrete choice models and limited dependent variable models in health economics. Finally, Section 6 summarizes and concludes by suggesting possible future lines of the development of the field.

## 2. SPATIAL DISCRETE CHOICE MODELS BASED ON CLASSICAL INFERENCE

In the three following subsections we are going to review spatial discrete choice models based on classical inference and their related problems. In particular, Section 2.1 will be on binary choice models, Section 2.2 on ordered choice models, Section 2.3 on multinomial choice models, Section 2.4 on count data models.

## 2.1. SPATIAL BINARY LOGIT/PROBIT MODELS: AN OVERVIEW OF DIFFERENT PROPOSED ESTIMATORS TO AVOID THE ESTIMATION PROBLEM IN SPATIAL MODELS

In econometric literature, a qualitative (or discrete) dependent variable can be differently described depending on if it assumes two or more modalities. In the simple case of due modalities, the dependent variable is defined binary or dichotomous. The first attempt to describe these variables was the *Linear Probability Model (LPM)* (*e.g.* Greene, 2003). Despite its simple use and interpretation of the parameters, as well as the fact that if the model contains a dummy variable for membership in some group, and every member of the group has the same value for the dependent variable, the coefficient of the group dummy variable can be only estimated in the linear probability model (*e.g.* Evans and Smith, 2005; Caudill, 1988), that model has gradually given up because of its main drawback: the possibility of having probabilities outside the (0,1) range, which causes inconsistency of the usual estimators. As a consequence, the linear probability model fails to take into account the dichotomous nature of the dependent variable. For that reason, the most utilized models are the *Binary Probit/Logit (BP/BL) models* (*e.g.* Greene, 2003; Verbeek, 2004), which are nonlinear models that take into account the binary nature of the dependent variable, and parametric models because they assume the error terms are distributed as a normal and a logistic distribution respectively. They are also known as *latent variable models* because of the highlighting of a continuous latent (*i.e.* unobserved) variable, which can be observed by an indicator binary variable. Both on methodological level and empirical level, particular attention has been paid on the **Spatial Autoregressive Probit Models (SAPM)** and S**patial Error Probit Models (SEPM)** (*e.g.* Fleming, 2004; Beron and Vijverberg, 2004; Holloway and Lapar, 2007), because the error term in the spatial logit models (*i.e.* SALM, SELM) is analytically intractable (Anselin, 2002). Beron and Vijverberg (2004) performed a set of Monte Carlo simulations of a **Spatial Linear Probability Model (SLPM)** compared with standard and spatial probit models. They found that, although the spatial linear probability model is much easier to estimate in terms of computational times than the spatial probit model, the first one fails to take into account the dichotomous nature of the dependent variable and it is not able to capture the spatial dependence in a theoretically adequately way. The standard probit model is able to capture the binary nature but it ignores spatial structure. Therefore, they concluded that spatial probit models are superior respect to the spatial linear models, and these last ones will become obsolete *as accessibility to spatial probit software becomes widespread*. Fleming (2004) offered a good overview on spatial discrete choice estimators and their main problems. The most important is that in Spatial Discrete Choice Models the spatial dependent structure adds complexity in the estimation of the parameters. In particular, one has to deal with two main problems: ***inconsistency*** *of the standard probit model because of the heteroskedasticity*

*introduced by spatial dependence* and **loss of efficiency** *of not using off-diagonal information (i.e. the correlation information) of the variance-covariance matrix*. If spatial dependence introduces heteroskedasticity, maximum likelihood (ML) estimates remain consistent assuming independent errors, but no longer efficient. In return, the joint probability function is reduced to the simpler product of independent density functions, that avoid the so-called *multidimensional integration problem* (*see* subsection b of this paragraph). In order to illustrate in details, in the following subsections we distinguish between these different problems and, for each of them, we show the relative proposed solutions, following more or less the same scheme proposed by Fleming (2004). In particular, subsection a presents solutions for the heteroskedasticity problem, subsection b shows solutions for the loss of efficiency due to the nonuse of the off-diagonal information, subsection c proposes solutions to avoid the computational burdensomeness of inverting $n$ by $n$ matrices, subsection d defines the spatial model as a weighted nonlinear version of the linear probability model, subsection e introduces the locally weighted regressions as alternatives to define spatial models, and finally subsection f introduces the social interaction and social network effects in Spatial Discrete Choice Models.

### a. *Avoiding inconsistency: solutions for heteroskedasticity*

A solution for heteroskedasticity in a Spatial Error Probit Model (SEPM) was first proposed by Case (1992) with the definition of a particular W matrix that accounts for heteroskedasticity and, therefore, the use of a heteroskedastic consistent maximum likelihood estimator. He allowed the spatial dependence using a structure that generates correlation among the observations within a region, but assumes that there is no correlation among the same observations in different regions. For the same model, particularly interesting was the work of Pinkse and Slade (1998) in a generalized method of moments (GMM) estimation framework. Their *GMM technique* is based on the moment conditions implied by the likelihood function for a spatial probit model with heteroskedasticity. They noted that the score vector (*i.e.* first-order derivatives of the log-likelihood function) can be viewed as a set of moment conditions which can be used with a GMM estimator. As in all GMM techniques, the existence of appropriate instruments is needed. They also stressed the need of more instruments than the exactly necessary ones to avoid identification problem. Unfortunately, drawbacks are not few. Firstly, parameters estimates may be questionable because the model relies on both large sample asymptotic properties for consistency of the estimates and asymptotic normality of the GMM estimator. Regularity conditions for consistency are based on the increasing demain asymptotics that seems to be a plausible assumption for lattice based data but not for micro level data (*see e.g.* Fleming, 2004 for details), as it is frequently the case in Health Economics. The asymptotic normality assumption is based on the condition that the dependence dies as distance increasing. These asymptotic conditions force researchers to pay attention on the choice of the weight matrix, because not all of them can satisfy these assumptions[3]. Moreover, parameter estimates will of course be sensitive to the choice of the particular structure of the weighting matrix (Robinson, 2008; Bivand, 2008; Bell and Blockstael, 2000). Secondly, parameters are *estimated simultaneously*[4] so that inverses of *n* by *n* matrices have to be calculated and an optimization complexity of the moment conditions for moderate-large samples makes practical application more difficult (*see* subsection c of this paragraph). Therefore, even if they did not use a ML estimation, the GMM estimator used in a Spatial Probit context is affected by the same problem of the ML estimator: the need of inverting *n* by *n* matrices which makes the previous methods computationally infeasible. Recently Pinkse *et al.* (2006) have proposed a *one-step GMM or continuous-updating (CU) estimator* (Hansen *et al.*, 1996) for a dynamic spatial (*i.e.* time-space) discrete choice model with fixed effects, endogenous regressors and arbitrary patterns of spatial and time-series dependence to investigate closing and reopening decisions for a panel of Canadian copper mines. They established asymptotic properties, *i.e.* consistency and asymptotic normality, of a different GMM-type estimator – CU estimator – which is a member of the class of *generalized empirical likelihood (GEL) estimators*[5]. Like standard GMM estimators, GEL estimators use moment conditions, but in over-identified models the statistical properties of GEL estimators tend to be superior in small and moderate sample sizes. Therefore the CUE is similar to the regular two-step GMM estimator, albeit that the weight matrix is parameterized immediately. They accounted for heteroskedasticity and endogeneity of some regressors, since for the latter problem the classical instrumental variable (IV) method cannot be applied. Moreover, they did not assume stationarity (*i.e.* the joint distribution can depend on location and not just on distance between locations) and they considered the increasing demain asymptotic problem by utilizing a new central-limit theorem (CLT) (Pinkse *et al.*, 2007). Finally, instead of including fixed effects in the latent variable equation, as it is usually done in the literature, fixed effects were linearly introduced in the observed-choice equation and so they could be removed by differencing, although the interpretation of their fixed effect is different. Comparing different models, their CU estimator works well on simple models. However, the authors also stressed that for more complex models the estimated coefficients became less significant. In a related work, Iglesias and Phillips (2008) have stressed that the second order finite properties of GMM and GEL estimators under weak dependence assumptions and nonstationarity are still unknown. Therefore, they showed how the asymptotic biases change in the weaker dependence and nonstationary setting, and they examined the need for a bias correction mechanism such as the bootstrap method.

---

[3] The **problem of choosing an appropriate W weighting matrix** is one of the most controversy discussion in the literature about the spatial parametric models. The arbitrariness of the spatial weight matrix $W$ is based on the fact that we usually do not know the actual spatial structure (Klier and McMillen, 2008). Many authors have stressed the importance of choosing an "appropriate" spatial weight matrix $W$, among whom for example Anselin (2010), Bivand (2008), Bell and Dalton (2007), Anselin (2002), Bell and Blockstael (2000). In particular, Bell and Dalton (2007) highlighted the problem of specifying a weighting matrix for micro-scale or individual data, in which the difficulties are related to correctly describe all the relationships among economic agents.

[4] We generally deal with the so-called **simultaneous equation models** because of the endogeneity of some regressors in spatial models. As a consequence, we need to estimate all the parameters at the same time. In the case of ML estimators this leads to a joint probability approach in order to specify Full ML (FML) functions.

[5] For an excellent historical and technical overview of different kind of econometric estimators, without the consideration of spatial effects, see Bera and Bilias (2002).

b. *Avoiding loss of efficiency: the use of the off-diagonal information and the related multidimensional integration problem*

If one wants both to address the heteroskedasticity and to use the off-diagonal correlation information of the variance-covariance matrix a *multidimensional integration problem*[6] arises, and computational techniques need to be used. In practical cases, the likelihood functions associated to the spatial autoregressive models cannot be analytically maximized due to the high degree of nonlinearity in the parameters. Therefore, this leads to the class of **Simulated ML (SML) estimators**, in which the log-likelihood function can be numerically maximized, but the computational procedures currently available in the software are all approximated. McMillen (1992) proposed the *expectation-maximization (EM) algorithm*, an estimation procedure that indirectly solves the multidimensional likelihood function in two steps and it is based on the principle that a possible outcome of the latent variable can be determined. The E-step (*i.e.* the expectation step) takes the expectation of the likelihood function for the latent variable conditional on the observed variable and a starting value for the parameter vector. The M-step (*i.e.* the maximization step) maximizes the resulting expected likelihood function for the parameter vector. These two steps are then repeated until the parameter vector converges. Since spatial parameter standard errors need to be estimated from the intractable $n$-dimensional dependence structure, no parameter standard errors are provided. Another drawback concerns the existence of a computational high cost in the repetitions of the algorithm since it requires calculation of the determinant of $n$ by $n$ matrices until convergence is assured, a problem strictly related to subsection c. Moreover, the author underlined that SAPM and SEMP are limited to small data sets since Limited Dependent Variable Models generally require large sample sizes before asymptotic results are accurate. In a related work, McMillen (1995) also showed that with smaller-size samples it is difficult to reject the homoscedastic probit model. Furthermore, even if a standard probit model is rejected, test statistics are not all clear in choosing between the SAPM and the SEPM (*see* Beron and Vijverberg, 2004). Beron *et al.* (2003) and Beron and Vijverberg (2004) proposed the *recursive importance sampling (RIS) simulator* (*i.e.* a generalization of the GHK simulator; *see e.g.* Bolduc *et al.*, 1997). This simulation method directly deals with the *n*-dimensional integration and it is based on the principle that it is possible to build a probability distribution that reflects the positive and negative errors around the mean, so that one can obtain estimates of the likelihood function that are closed to the actual likelihood value. The authors also stressed that this technique can be used for studies like that of Murdoch *et al.* (2003), who used a full maximum likelihood (FML) estimator. Despite its advantages of providing unbiased spatial parameter standard errors and solving directly the high-dimensional integration as well as taking heteroskedasticity into account, its main drawback is the computational burdensomeness, which is the highest respect to the others computational techniques (*see* Fleming, 2004). In recent years, due to the possible computational advantages over the ML estimator, many researchers are increasingly incline to use Bayesian inference and its computational algorithms in Spatial Discrete Choice Models (*e.g.* LeSage, 2000; Holloway *et al.*, 2002; Rathbun and Fei, 2006; among many others). One of them is the so-called *Gibbs Sampling* (*e.g.* LeSage, 2000), whose analysis is referred to paragraph 3 dedicated to Bayesian Spatial Discrete Choice Models.

---

[6] The **multidimensional integration problem** is related to the fact that the $n$-dimensional likelihood function is analytically intractable since it is not possible to simplify the joint function into the product of the marginals as in the case of independent observations (*see e.g.* Fleming, 2004; Arbia and Billé, forthcoming).

### c. *Avoiding inverse of n by n matrices: methodological solutions*

When we specify a spatial autoregressive probit model we generally derive the maximum likelihood function implied by the reduced form of the spatial model (Arbia and Billé, forthcoming). A major problem in maximizing the log-likelihood function implied by the *reduced form*[7] of the spatial model is represented by fact that it repeatedly involves the calculation of the determinant of a *n* by *n* matrix (related to the weight matrix) whose dimension depends on the sample size[8]. When *n* is very large, this operation can be highly demanding even with the current computational power, as it often happens in many empirical application field like that of Health Economics. For this purpose many solutions have been proposed (Ord, 1975; Griffith, 2000; Smirnov and Anselin, 2001; Pace and LeSage, 2004). Moreover, the problem becomes unmanageable when dealing with very large and strongly connected matrices, as are those frequently encountered in social interaction applications (*see e.g.* Brock and Durlauf, 2007; *see* subsection f) and in applications with microlevel data. Unfortunately, for the last case it seems we are still not able to identify a feasible weighting matrix (*see e.g.* Bell and Dalton, 2007; Bell and Bockstael, 2000). This problem is important because it precludes the opportunity of making studies on a large scale, in a way that a comparison between more detailed spatial units becomes impracticable. Recently, based on the work of Pinkse and Slade (1998), Klier and McMillen (2008) have proposed a *linearized version of Pinkse and Slade's GMM* to avoid the infeasible problem for large samples of inverting $n$ by $n$ matrices, that also affected the simulated ML procedures (McMillen, 1992; Beron and Vijverberg, 2003, 2004), and the simulated Bayesian approach (LeSage, 2000). In fact, in that case no matrix needs to be inverted and estimation requires only standard probit/logit models and linear two stage least squares (2SLS). Linearization allows the model to be estimated in two steps. The first step is a standard logit model, in which spatial autocorrelation and heteroskedasticity are ignored. The second step involves 2SLS estimates of the linearized model. Therefore, standard GMM reduces to non-linear 2SLS. Monte Carlo experiments suggest that the linearized model accurately identifies spatial effects. In a study of auto supplier plants in the United States, a significant positive spatial autocorrelation was found. They performed a Spatial Error Logit Model (SELM) which can readily extend to the Spatial Autoregressive Logit Model (SALM), as well as SAPM and SEPM, although they stressed that the assumption in which the latent variable depends on its spatially lagged values may be disputable. In addition to avoiding inverse matrices, the choice of using a GMM approach is also due to the fact that this method does not impose a functional form of the error terms as in the ML case. This leads the GMM into the class of semiparametric estimators. On the other hand, the primary advantage of the maximum likelihood estimation, as they stressed, is the potential for efficiency[9] (*see also* Wang *et al.*, 2013), although the choice of $W$ is *arbitrary* and so the prospect of efficiency may become questionable when the true model is uncertain. For the last reason, they also stressed that GMM estimators are more robust than ML estimators. Finally, although their linearized version provides accurate estimates as long as the spatial coefficient is small, in comparison with the standard GMM the linearized version has higher standard errors, that is there is a loss of efficiency. Of different view was the work of Wang *et al.* (2013). They proposed a *Partial Maximum Likelihood Estimation (Partial MLE)* of a Spatial Error Probit Model (SEPM) with cross-section data. In their study they captured spatial dependence by considering spatial sites to form a countable lattice, and explore a middle-ground approach which trades off between efficiency and computational burdensomeness. The basic idea was to divide observations into many small groups (*i.e.* clusters) in which adjacent observations belonged to one group (*i.e.* pairwise groups), and bivariate normal distributions were specified within each group. Correctly specifying the conditional joint distribution within groups (*i.e.* utilizing relatively more information of spatial correlations), they estimated the model by partial maximum likelihood estimation which is consistent, asymptotically normal and more efficient than the GMM estimator of Pinkse and Slade (1998), because it uses both diagonal and off-diagonal correlations information between two closest neighbors. In fact, Pinkse and Slade (1998) did not taken advantage of information from spatial correlations among observation, using heterogeneities of the diagonal terms of the variance-covariance matrix, since they only consider the problem of heteroskedasticity. Wang *et al.*'s (2013) approach is subject on bias variance-covariance matrix estimators and, therefore, they followed Conley's (1999) approach to get consistent variance-covariance matrix estimators. Obviously, their estimator is not as efficient as the Full ML (FML) estimator. However, since information from adjacent observations usually captures relevant spatial correlations in the whole sample, they have provided a consistent and "relatively" efficient estimator, which avoid computational problem at expense of a loss of a relatively small part of efficiency. A last relevant paper on the estimation problem for spatial discrete choice models is that of Bhat and Sener (2009). Instead of trying to find a computational solution (*e.g.* McMillen, 1992; Beron and Vijverberg, 2004; LeSage, 2000), they proposed a *copula-based approach* in a spatial correlated heteroskedastic binary logit (SCHBL) model to accommodate spatial dependence between logistic error terms using a multivariate logistic distribution based on the Farlie-Gumbel-Morgenstein (FGM) copula, while also controlling for heteroskedasticity. Since the resulting spatial logit model retains a simple closed-form expression, no simulation machinery is involved, leading to substantial computation gains relative to current methods to address spatial correlation. Furthermore, it can be estimated using *standard and direct maximum likelihood estimation* and it is computationally tractable even for large sample sizes. Since closed-form techniques are more accurate than simulation techniques, this methodology invites researcher to formulate closed-form models rather than simulation-based models. A copula approach basically involves the generation of multivariate joint distribution, given the marginal distribution of the correlated variables. Therefore, a copula is a function that generates a stochastic dependence relationship among random variables with pre-specified marginal distribution. Based on the FGM copula family, they developed a particular multivariate variant of the Gumbel Type III bivariate logistic distribution. One limit of their spatial logit approach is that the correlation between observations is limited to moderate levels. However, the authors stressed that the correlation range of the FGM logistic distribution is not likely to be too limiting, since the correlation between observational units drops off sharply with geographic distance. Since their empirical application is focused on Health field, the empirical analysis is referred to paragraph 5.

---

[7] Due to the simultaneity of these models and the impossibility of observing the latent continuous variables, an easy analytical solution is precluded (*e.g.* Anselin, 2002).
[8] That is $|I_n - \rho W_n|$ for a Spatial Autoregressive model.

---

[9] One has to keep in mind that if the normality assumption is correct, semiparametric estimators are less efficient than the relative parametric ones.

### d. *Spatial Discrete Choice Models as weighted non-linear version of the linear probability model: the case of the spatial parameter treated as a nuisance parameter*

A different solution to avoid problems in Spatial Discrete Choice Models is to describe spatial discrete choice problem as a *weighted non-linear version of the linear probability model* with a general variance-covariance matrix that can be estimated by GMM (Hansen, 1982; Fleming, 2004). The consideration of a *non-linear weighted least squares (NLWLS) estimation* eliminates both calculation of $n$ by $n$ determinants and multidimensional integration problem of the Full ML approach, since no formulation of the maximum likelihood function is done. Therefore, this procedure is computationally simple even in large samples. The estimators thus derived are consistent and asymptotically normal, but in small samples they can be biased and they are not fully efficient. For a SAPM specification this estimator is a *weighted non-linear form of the two stage least squares or instrumental variable (WNL2SLS-IV) estimator*. The endogenous spatially lagged dependent variable is treated as any non-spatial endogenous variable in GMM framework (*e.g.* Kelejian and Prucha, 1998) where the ideal set of instruments are the increasing in order linear combinations of the exogenous variables and spatial weights matrix. For the SEPM specification this estimator is a *weighted non-linear form of the feasible generalized least squares (WNLFGLS) estimator*. Fleming (2004) proposed a GMM estimation that differs from that of Kelejian and Prucha (1999) in that the linear model is replaced by a non-linear model, constructing a non-linear least squares estimator based on three moment conditions. Even when the assumption of no spatial correlation is not correct, this estimator remains consistent although less efficient. These estimators minimize moments equivalent to the probit log-likelihood score vector when errors are *iid* and no spatial autocorrelation exists. In fact, a maximum-likelihood probit estimator with no autocorrelation is equivalent to a non-linear weighted least squares (NLWLS) estimator (McMillen, 1992). The main drawback of the GMM estimator for a Spatial Error Probit Model (SEPM) is that the significance of the spatial parameter cannot be assessed since it is considered a *nuisance parameter* (*i.e.* it is not considered an information parameter, so that the spatial correlation is *not substantive* as in a SAPM case; *see also* Anselin, 2002) that must be accounted for to improve the efficiency of regression coefficients and consistency of standard errors. In this case spatial standard errors are not provided. It is a controversy discussion the appropriateness of considering a spatial parameter only a nuisance parameter as also Beron and Vijverberg (2004) have stressed.

### e. *Locally weighted regressions and nonparametric methods*

Spatial econometric methods help to account for the effects of missing variables that are correlated over space. Although the use of a spatial contiguity matrix is usually the starting point to specify the relationship between neighboring observations, it has the disadvantage of imposing restrictive structure that can bias results when inappropriate (McMillen and McDonald, 2004). For that reason, *locally weighted regressions (LWR) or geographically weighted regressions (GWR)* (*e.g.* Fotheringham *et al.*, 1998; McMillen, 1992), as well as other *nonparametric estimation techniques*, are becoming prevalent also in discrete choice setting. The GWR/LWR and the Expansion Method (EM) for spatial data are two statistical techniques which can be used to examine the spatial variability (*i.e. spatial heterogeneity*) of the regression results across a region and so inform on the presence of spatial nonstationarity. The basic idea is that simple econometric models represent the data best in small geographic areas. In estimating separate functions for several spatial units, we are recognizing that their structure is sufficiently different such that the data should not be pooled. In other words, rather than accept one set of "global" regression results (*i.e.* the simultaneous autoregressive (SAR) model), both techniques allow the possibility of producing "local" regression results by specifying each parameter of the global regression as a function of the spatial coordinates with the use of linear expansions. The GWR/LWR is readily adaptable to discrete dependent variable models in a way that it constructs separate estimates for each observation with more weight given to nearby sites. Locally weighted estimates for a single observation is simply obtained by weighted least squares (WLS) (McMillen and McDonald, 2004). This methodology captures the essential idea behind Spatial Econometrics – that nearby observations are more closely correlated than those farther away – without imposing an arbitrary parametric weighted scheme. Furthermore, limiting the estimation to a neighborhood of an observation while allowing for nonlinearity eliminates much of the heteroskedasticity and autocorrelation in spatial data.

### f. *Social interactions and social network effects*

In recent years, it became more common to include social interactions or neighborhood effects in Spatial Discrete Choice Models (*i.e. social network effects*), especially in transportation modeling

(*e.g.* Goetzke and Andrade, 2010; Páez *et al.*, 2008a; Páez *et al.*, 2008b; Goetzke, 2008; Páez and Scott, 2007; Brock and Durlauf, 2007). In particular, Goetzke and Andrade (2010) stressed that it is important to include social interactions and correlated effects in mode choice models as one combined spatial spillover variable for two reasons: spatial spillover serves the purpose to avoid a possible omitted variable bias, and, in addition, the spatial spillover variable can be seen as a proxy for the mode-friendliness in the neighborhood. Within the context of choice to walk, their study focused on investigating the spatially autoregressive structure in a binary mode choice modeling to describe the choice of walking in New York City. They proposed an *instrumental variable (IV) approach* for estimating the spatial lag coefficient, in conjunction with a linear probability mode choice model and a logit mode choice model. For the linear probability model the IV approach is essentially a 2SLS estimation. In that case a correction for heteroskedasticity is also provided with the weighted least squares (WLS) method. They found in both cases that the walkability variable or instrumental variable (*i.e.* combination of the social interactions with correlated effects) improves the regression, and moreover, it avoids an omitted variable bias and it is significantly positive. However, the linear probability model is restricted to just two choices, while the logit specification can be applied to McFadden-type conditional mode choice models as well as multinomial choice models (*e.g.* Páez and Scott, 2007; Páez *et al.*, 2008b; *see* paragraph 2.3).

## 2.2. SPATIAL ORDERED PROBIT/LOGIT MODELS

When the dependent variable assumes more than two outcomes ordered between them, *Ordered Probit (OP) models* are usually adopted (*e.g.* Greene, 2003; Verbeek, 2004), and they are often specified by unknown thresholds (*e.g.* Jones, 2000). Recently, Munkin and Trivedi (2008) proposed a Bayesian Ordered Probit model with Endogenous Selection (Bayesian OPES) to study the effect of different types of medical insurance plans on the hospital utilization level in a UK population aged between 55 and 75 years. The authors have also highlighted that this model can be used for count data (*see* paragraph 2.4), because it analyzes the effects of a set of endogenous choice indicators on a count variable whose distribution has a high percentage of zeros. Although many databases require ordered discrete responses in a spatial context, no much papers were found. Some spatial ordered probit models were proposed in a Bayesian framework and in Health Economics as in the previous example, so their analyses are referred to paragraph 3 and 5 respectively.

## 2.3. SPATIAL MULTINOMIAL PROBIT/LOGIT MODELS

When a discrete dependent variable assumes more than two modalities that cannot be ordered between them, the appropriate models are known as the *Multinomial Probit/Logit (MNP/MNL) models* (*see* Greene, 2003; Verbeek, 2004), which are generally justified by the **random utility theory** (*e.g.* Greene, 2003; McFadden and Train, 2000; Dowd *et al.*, 1991; McFadden, 1974). Multinomial logit (MNL) models define the individual utility function based on some features that only vary between individuals (*i.e. effects through decision-makers*) and some others that only vary among individual choices (*i.e. effects among choice alternatives*). This class of models is essentially based on two strong assumptions: (i) the error terms are *independent and identically distributed (iid)* with a type I extreme value (or Gumbel) distribution, (ii) the existence of the *unobserved response homogeneity*. Their combination constitutes an analytical advantage because the individual choice probability distributions are in a closed-form, but at the same time it is also considered its main drawback because it implies the so-called *independent of irrelevant alternatives (IIA) property* (*e.g.* Bolduc, 1992; Bolduc *et al.*, 1996a; Bhat and Guo, 2004). Since this property reflects an individual choice independence, the restrictiveness of the IIA property for several applications, including those of the Health field (*e.g.* Foster *et al.*, 1996; Akin *et al.*, 1995; Feldman *et al.*, 1989), has been highlighted by many researches. From a probability point of view, this assumption consists in having probabilities that *proportionally decrease* as a new alternative is introduced in the set of alternatives (*e.g.* Jones, 2000; Foster *et al.*, 1996). Alternatively, we can say that the ratio of the choice probabilities for two alternatives is unaffected by the presence of any other alternative. Therefore, a change of the probability of one alternative will lead to identical changes in the relative choice probabilities for all the other alternatives, that is the cross-elasticities are constant (*e.g.* Hunt *et al.*, 2004). Obviously, the IIA is unlike to hold in spatial choice applications as well as in many Health applications, so that the model was progressively generalized. Depending on which of these two assumptions are relaxed, (i) or (ii), we can formulate a different multinomial logit model. All of these variants belong to the class of the *Mixed Logit (ML) models* (*see e.g.* Koppelman and Sethi, 2005 for an excellent overview of the origin of different random utility models). As a consequence, Greene and Hensher (2003) have proposed a semi-parametric extension of the MNL model based on a latent class formulation which relaxes the MNL requirement that the analyst makes specific assumptions about the distributions of parameters across individuals. Recently, particular attention has been paid on *Generalized Extreme-Value (GEV) models* (*e.g.* Hunt *et al.*, 2004; Bhat and Guo, 2004; Bekhor and Prashker, 2008; Pinjari, 2011), in which the independence assumption of the error terms *across alternatives* is relaxed, as well as on different *Multinomial Probit (MNP) models* (*e.g.* Bolduc, 1992; Greene, 2003), in which the error independent assumption does not imply the IIA property. Hunt *et al.* (2004) offered an excellent overview on spatial discrete choice developments, focusing on GEV models. In particular they analyzed a type of GEV model called *generalized nested logit (GNL) model* (Wen and Koppelman, 2001), pointing out its main assumptions and limits. One important drawback seems to lie into its utility maximization requirement that is frequently violated in many applications. In this case the model is no more consistent with random utility theory. Due to the rigid error correlation structure requested a priori by the GEV models, Bolduc (1992) sustained that MNP model was the best "potential" model to take alternative interdependences into account. To avoid the problem of *increasing nuisance parameters as alternatives increases*, he proposed a MNP with first-order generalized [GAR(1)] errors, providing also rank conditions for model identifiability. For a good comparison between MNL, independent MNP e MNP models see also Bolduc *et al.* (1996a). Garrido and Mahnassani (2000) employed a Spatial Dynamic Multinomial Probit (SDMNP) model to predict the freight pickup choices of a large truckload carrier in Texas. A Monte Carlo simulation was performed to evaluate the MNP likelihoods. They assumed a general error structure in which the error component in a given region is spatially correlated with the errors in other regions, and the error component is the outcome of a dynamic stochastic process during a given interval of time. Nelson *et al.* (2004) utilized three different models to econometrically estimate a spatially explicit economic model of a proposed road improvement activity in Panama's Darién province and they simulated location-specific effects on land use. They used a standard Multinomial Logit (MNL) model, a Nested Logit (NL) model, and a Random Parameters Logit (RPL) model (*i.e.* mixed MNL model, *see* McFadden and Train, 2000), highlighting that the last one is the most appropriate since the standard MNL is affected by IIA property and the NL does not take heteroskedasticity into account. They also used a coding technique (Besag, 1972c) to reduce potential spatial correlation by destroying the information of the underlying spatial process. Therefore, in that case the spatial parameter is treated as a nuisance parameter. Among GEV model supporters, Bhat and Guo (2004) have proposed a Mixed Spatially Correlated Logit (MSCL) model which utilized a GEV structure in order to consider utility correlation between spatial units, and it superimposed a mixed distribution on the GEV structure to capture the unobserved response heterogeneity. They employed a restricted form of the generalized nested logit of Wen and Koppelman (2001) to examine the housing choices of residents from Dallas County, Texas. In the same way, Bekhor and Prashker (2008) examined several GEV models to discuss their adaptability on destination choice situations, with the object to determine the probability that a person from a given origin chooses a particular destination among different available alternatives. Taking Bhat and Guo's work, they proposed a model with three hierarchical levels. Recently, Pinjari (2011) has formally obtained the class of Multiple Discrete-

Continuous Generalized Extreme Value (MDCGEV) models, and in particular he tested the existence and extracted the general form of the consumption probability in a closed-form, and he next applied the model in a household expenditure analysis. In the *social network context*, two important papers interconnected between them are those of Páez and Scott (2007) and Páez *et al.* (2008b). Their analysis dealt with the development of a multinomial logit model that includes both conventional factors as well as alternative features and some social aspects (*e.g.* externalities) which can influence interaction between economic agents in the space.

## 2.4. SPATIAL DISCRETE CHOICE MODELS FOR COUNT DATA

*Count data models* are used when the dependent variable consists in a count of positive integers (*e.g.* Greene, 2003; Verbeek, 2004). Due to the nature of these variables, the data are usually affected by *asymmetric distributions* and a *high portion of zeros*. The basic model for count data is the *Poisson model* (also known as the "law of rare events"). However, this model has limited applications in Health Economics because of its *equidispersion property* (*i.e.* the mean equals the variance), which does not usually arise in many dataset. For example, studies on *Health care utilization* (*e.g.* Cauley, 1987; Mullahy, 1997b; Pohlmeier and Ulrich, 1995; Gurmu, 1997 among many others) often showed evidences of overdispersion, in which case the Poisson model underestimated the actual zero frequency and the right-tail values of the distribution. *Negative Binomial (NB) models* are usually adopted to avoid the equidispersion problem (*e.g.* Cameron and Johansson, 1997; Geil *et al.*, 1997; Gerdtham, 1997; Cameron and Trivedi, 1986, among many others). Despite Mullahy (1997b) stressed that the presence of zeros is "a strict consequence of the *unobserved heterogeneity*", this presence can also occur without unobserved heterogeneity, in which case it is necessary to use either *zero-inflated (i.e. with zero/ZI) models* (*e.g.* Mullahy, 1986) or *hurdle (i.e. two-part/TP) models*. The former consist of a mixed formulation which give more weight to the probability of observing a zero. The latter is characterized by two independent probability processes, which are generally a binary process and a Poisson process. Moreover, a *modified hurdle model (MHM)* was developed by Santos-Silva and Covas (2000) to account for underdispersion of the data. In more recent years, many discrete choice studies are seeing the diffusion of *finite mixture models (FMMs)* (*i.e. latent class models (LCMs)*) in which the random variable is supposed to be drawn from a mix of different distributions (*e.g.* Winkelmann, 2004; Deb and Trivedi, 2002). These kinds of models include the above-mentioned zero-inflated models. In the Health field, especially in Health care utilization studies, these models were compared to the hurdle models bringing a debate (*e.g.* Deb and Holmes, 2000; Jiménez-Martin et al, 2002; Winkelmann,

2004 among others). A substantial difference between these two classes of models is that TPMs distinguish between users and not users of the Health care services (*i.e.* because of the binary process), whereas FMMs distinguish among frequently users and not frequently ones. Bago d'Uva (2006) proposed a *finite mixture hurdle panel (FMH-Pan) model*, which is a model with both FMM and TPM. However, the author stressed that for cross-sectional data the model is characterized by identification problems. Finally, as an alternative to FMM, Machado and Santos-Silva (2002) developed a *quantile regression model* for count data, which is considered a promising one in order to solve the above-mentioned problems. Empirical econometric papers on count data models in a spatial setting are still not many. Some of them utilized a Bayesian estimation and/or are related to Health Economics, so their analysis is referred to paragraph 3 and paragraph 5 respectively.

## 3. BAYESIAN SPATIAL DISCRETE CHOICE MODELS

In recent years, we have experienced an increasing diffusion of *Bayesian inference and MCMC techniques* (*see e.g.* Chib and Jeliazkov, 2001; Geweke *et al.*, 1997; Chib, 1995; Chib and Greenberg, 1995; Albert and Chib, 1993; Casella and George, 1992; Chib, 1992; McCulloch and Rossi, 1991) into Spatial Econometrics science (*e.g.* Lacombe, 2008; Smith and LeSage, 2004; Hepple, 2003; LeSage, 2000, 1997). In a computational setting, the *Gibbs sampler* is similar to the EM algorithm in a way that solves indirectly the n-dimensional integration, formulating a likelihood function as if the dependent variable were continuous. However, it is different in the way it formulates the likelihood function and the estimates of the unobserved latent variable (Fleming, 2004). This method is also characterized by a data augmentation step that links the observed discrete dependent variable with the latent one. More important, this method overcomes the standard error problem of the EM algorithm because standard errors can be obtained from the posterior parameter distribution. This is the most flexible of the computational-based spatially dependent models because it can incorporate spatial lag dependence and spatial error dependence in addition to general heteroskedasticity of unknown form. The Gibbs sampling uses conditional posterior distributions to achieve estimates of the parameters in the unconditional posterior distribution. This technique was generalized by the *Metropolis-Hasting algorithm* if the conditional distribution has an unknown form. For that reason, computational Bayesian techniques may be more attractive than the ML-based ones. However, it is necessary to keep in mind some general aspects of the Bayesian inference. From a statistical point of view, one of the most important problems of the Bayesian inference is choosing a "correct" (*i.e.* proper) prior distribution, which reflects our knowledge of the parameters without any additional information. If a non-correct prior distribution is chosen, the estimates may give misleading results. Moreover, in most cases one may also not have this prior knowledge. Many uninformative or diffuse priors (also known as flat prior like the uniform distribution) have been proposed in the literature (*e.g.* Jeffreys' prior) (*see e.g.* Lee, 2004), in which cases we used a Bayesian estimation without having prior knowledge about the parameters. In other cases, if the sample size is very large this probably leads the case in which "*the likelihood dominates the prior*", so that the results can be the same as we only consider the maximum likelihood estimation. In conclusion, being Bayesian inference a different "way" to view the estimation of parameters, a comparison with ML and other kind of estimators is needed to make us sure of our results. For example Bolduc *et al.* (1997) conducted a comparison between Maximum Simulated Likelihood (MSL) estimation with the GHK simulator and the Gibbs sampling. No significant differences were found between them, although the Gibbs approach is conceptually and computationally simpler to implement. Analyzing empirical papers, in the case of *binary modalities*, Holloway *et al.* (2002) developed a Bayesian Spatial Autoregressive Probit Model (Bayesian SAPM) to estimate the autoregressive spatial parameter, the *neighborhood effects*, of the adoption of high-yielding variety (HYV) between rice makers in Bangladesh. Albers *et al.* (2008) also utilized a Bayesian SAPM in a study of how new public reserves influence the configuration of the private land conservation. The authors highlighted the absence in the literature about a test that is able to discriminate between SAPM and SEPM. Furthermore, they stressed there is not an estimator able to account for endogeneity of one or more regressors. For the former, one possible solution could be the developing of a SARAR(1,1) for probit/logit models. For the latter, one possible solution seems to be a GMM estimation for probit models (*e.g.* Pinkse and Slade, 1998; Klier and McMillen, 2008). Finally, Jaimovich (2010) presented a Bayesian heteroskedastic spatial probit model in a study about the Free Trade Agreement contagion.

In the case of *ordered modalities*, Wang and Kockelman (2009a,b) developed a Bayesian Dynamic Spatial-Ordered Probit (Bayesian DSOP) model in order to capture spatial and temporal autocorrelation patterns on ordered categorical data, and they applied it to assess patterns of land development change in Austin (Texas).

In the case of *non-ordered modalities* (*i.e.* MNL/MNP models), Wall and Liu (2009) developed a Spatial Latent Class Analysis (SLCA) model by adding a spatial structure to the distribution of the latent class (*i.e.* the discrete variable) with a MNP model. Chakir and Parent (2009) considered a Bayesian Spatial Multinomial Probit (Bayesian SMNP) model to identify the determinants in land use changes, at a level of single parcel (*e.g.* microeconomic level), in the Département du Rhones in France from 1992 to 2003. The model is based on the assumption that landholders can choose between four categories of land use for a given parcel and a given instant of time: (i) agricultural,

(ii) forester, (iii) urban, (iv) no use. Furthermore, the model takes both spatial dependence between closed parcels and interdependence among land use alternatives into account. The choice of a micro level analysis is justified by the fact that previous studies in that field were at a macro level analysis, due to the large availability and the low cost of the aggregated data. In fact, the authors highlighted that in macro level analysis you can usually lose the heterogeneity of observations that characterizes each region (*see e.g.* Anselin, 2002).

In the *count data case*, Rathbun and Fei (2006) developed a Bayesian Spatial zero-inflated Poisson (ZIP) model, *i.e.* a zero-inflated Poisson model in which the excess of zeros are generated by a spatial probit model (*see also* Heagerty and Lele, 1998), in order to analysis the regeneration of oaks (*i.e.* quercus species) using both physical and biotic variables and data from 38 mixed-oak stands surveyed in central Pennsylvania from 1996 to 2000. A spatial version of the probit model is obtained by letting the elements of the variance-covariance matrix to depend on the distance between pair of sites. The choice of the Bayesian approach was not justified by their prior knowledge, but it was adopted because of the lack of proves of the ML large-sample inferential properties. In fact, they highlighted that "little information is available for the elicitation of priors in the present paper", that is a proper prior to ensure a proper posterior distribution cannot be chosen. However, they expected that given the large sample size "the data could dominate the prior", that is the ML function provided the significant information for their study, without therefore any significant difference between Bayesian and ML approach. In cases like those of Rathbun and Fei (2006), it is unclear the choice of using a Bayesian approach without having any substantive reason. Ver Hoef and Jansen (2007) proposed Bayesian space-time zero-inflated Poisson (ZIP) and Hurdle models to investigate haulout patterns of harbor seals on glacial ice in Disenchantment Bay, Alaska. This models have been constructed by using spatial conditional autoregressive (CAR) models and temporal first-order autoregressive [AR(1)] models as random effects.

Finally, in a *limited dependent variable context*, Mizobuchi (2005) developed a Bayesian Spatial Tobit (Bayesian ST) model to estimate the "adjacent effect" in the electrical installation behaviors in a study on the $SO_2$ authorization market.

## 4. LIMITED DEPENDENT VARIABLE MODELS

Although a lot of papers of Limited Dependent Variable Models with an application in the Health field were found, because of the peculiarities and the micro-scale of health data (*see* paragraph 5), Spatial Limited Dependent Variable Models are not so diffuse as well as aspatial ones. In particular, to the best of our knowledge, any spatial limited dependent variable model was found. That said, since the large diffusion in Health Economics of Limited Dependent Variable Models with no spatial autocorrelation, a brief description of these types of models is necessary.

In order to describe Limited Dependent Variable Models we need to distinguish between censored data, truncated data and duration data (*see e.g.* Greene, 2003; Verbeek, 2004). In particular, we exclude from this analysis the models related to the last type of data, *i.e.* the *duration models*.

In the case of *censored data*, the typical models are the *Censored Regression Models or Tobit I Models* (Tobin, 1958). In these cases the observed dependent variable is continuous[10], but it is observed only for a part, usually positive, of the overall distribution. Therefore, the main characteristic of the limited dependent variables is the *high skewedness* of their distributions. More precisely, the most frequently case is that the observed dependent variable is null for negative values of the continuous latent dependent variable and it is positive for positive values of the continuous latent dependent variable, so that the negative values of the latent variable are all changed in zero in the observed variable. In other words, the observations are lower down *censored in zero*. In the case of *truncated data*, the sample is drawn from a subset or restricted part of a larger population of interest. In particular, if the form of truncation is an *incidental truncation*, a sample selection problem arises, so that the sample under observation is no more a random sample. Typical models that overcome this problem are the *Sample Selection Models (SSM) or Tobit II Models*. Differently from the censored case, truncated data are those in which the observations associated with the negative values of the continuous latent variable are completely excluded from the sample. In other words, for negative values of the latent variable the observed dependent variable is *not observed*, and it is not set to zero as in the censored case. A class of applications in which the selection problem plays an important role is the **Treatment Effects Analysis**. Most of the papers in Health Economics were published in a study on treatment effects (*see* paragraph 5).

The previous class of models can be viewed as *Two-part models (2PM/TPM)*, that is to say they are models composed by two parts. The first part describes a binary choice problem, whereas the second part describes the distribution of the phenomena under study. Applications of the TPM in Health Economics have usually used the logarithmic transformation to avoid the skewedness of the distributions. Some others have preferred the use of the squared root (*e.g.* Veazie *et al.*, 2003) and the Box-Cox transformation (*e.g.* Chaze, 2005). For details one can see Jones (2010). However, in order to draw some political conclusions on the estimated parameters a *problem of retransformation* arises. This problem leads the researchers to use *nonlinear models* like for example the *Exponential Conditional Mean (ECM) models* (*e.g.* Basu and Manning, 2006) and the class of Generalized Linear Models (GLMs). Recently, more flexible extensions of the GLMs like the *Generalized*

---

[10] This is the main difference between limited dependent variable models and discrete choice models.

*Gamma Models (GGMs)* (*e.g.* Manning *et al.*, 2005) and the *Extending Estimating Equation (EEE)* approach (*e.g.* Basu and Rathouz, 2005) have been proposed.

## 5. SPATIAL DISCRETE CHOICE MODELS AND LIMITED DEPENDENT VARIABLE MODELS IN HEALTH ECONOMICS

In the last 20 years, we have had an increased experience in econometric studies as basis of health policy. In particular, Newhouse (1987) highlighted the need of using applied econometric techniques also in Health Economics for two main reasons. The former is that Health Economics is an applied field with great political interest. Newhouse stressed the fact that at least 10 percent of the GNP (Gross National Product) was spent in health care, and public programs accounted for about 40 percent of American expenditure on personal health care. For a description of the *Italian Healthcare system* see for example France *et al.* (2005). The latter is that one of the most crucial point concerns the Health Care Expenditure, whose distribution across individuals is very skewed. In fact, the prevalence of latent variables, unobserved heterogeneity, nonlinear models, and other peculiar characteristics, makes Health Economics a particular rich area of Applied Econometrics (Jones, 2000).

As a consequence, *Health Econometrics* is still a developing field. A recent review on statistical methods for analyzing Healthcare resources and costs, particularly focused on the use of experimental data, was that of Mihaylova *et al.* (2010). They defined Health Econometrics as "*a field characterized by the use of large quantities of mostly observational data to model individual healthcare expenditures, with a view to understanding how the characteristics of the individual, including their health status or recent medical experience, influence overall costs*". In addition, they also argued that "*observational data are vulnerable to biases in estimating effects due to non-random selection and confounding that are avoided in randomized experimental data*".

According to whether we analyze individual data or aggregated data, we can apply the microeconometric and macroeconometric techniques respectively, but in most cases the above-mentioned peculiarities of health data make us unable to correctly use these econometric methodologies. As a consequence, the econometric techniques and the utilized models in Health Econometrics differ according to different observed data and they are continuously subject to criticisms and improvements by researchers. As a matter of fact, in the last thirty years there has been in Health Economics literature a vigorous debate about the most appropriate econometric model to be used between the Two-part model (2PM or TPM) and the Sample Selection Model (SSM/Tobit II) to describe the *Demand for Medical Care* and the *Health Care Expenditure* (*see e.g.*

Madden, 2008; Norton *et al.*, 2008; Jones, 2000; Hay *et al.*, 1987; Duan *et al.*, 1984; among many others).

In Health Economics the most utilized models are usually Discrete Choice Models (*i.e.* binary, ordered, multinomial, count data models) and Limited Dependent Variable Models (Tobit I, 2PM and SSM/Tobit II, Duration models) (*e.g.* Greene, 2003; Verbeek, 2004). For an overview of the different models utilized in Health Economics see Jones (2000). Many papers have been written without the consideration of the spatial parameters. Some of them are the following:

- ❖ **Discrete choice models:**

  - ➢ **Linear probability models** (*e.g.* Evans and Smith, 2005 among others)

  - ➢ **Binary probit/logit models** (*e.g.* Nketiah-Amponsah, 2009; Wolff and Maliki, 2008; Iram and Butt, 2008; Ai and Norton, 2003 among many others)

  - ➢ **Ordered probit/logit models** (*e.g.* Varin and Czado, 2010; Munkin and Trivedi, 2008; Contoyannis *et al.*, 2004; Lindeboom and van Dooslaer, 2004; van Dooslaer and Jones, 2003; Groot, 2000 among many others)

  - ➢ **Multinomial probit/logit models** (*e.g.* Lindeboom and Kerkhofs, 2009; Bolduc *et al.*, 1996a; Foster *et al.*, 1996; Akin *et al.*, 1995; Mwabu *et al.*, 1993; Dowd *et al.*, 1991; Feldman *et al.*, 1989 among many others)

  - ➢ **Count data models** (*e.g.* Bago d'Uva, 2006; Winkelmann, 2004; Deb and Trivedi, 2002; Jiménez-Martín *et al.*, 2002; Santos Silva and Windmeijer, 2001; Deb and Holmes, 2000; Santos Silva and Covas, 2000; Munkin and Trivedi, 1999; Windmeijer and Santo Silva, 1997; Gurmu, 1997; Pohlmeier and Ulrich, 1995 among many others)

- ❖ **Limited dependent variable models:**

  - ➢ **Two-part models** (*e.g.* Madden, 2008; Norton *et al.*, 2008; Stuart *et al.*, 2008; Tiang and Huang, 2007; Welsh and Zhou, 2006; Deb *et al.*, 2006; Basu *et al.*, 2004; Buntin and Zaslavsky, 2004; Veazie *et al.*, 2003; Manning and Mullahy, 2001; Goldman *et al.*, 1998; among many others)

  - ➢ **Sample selection models** (*e.g.* Madden, 2008; Norton *et al.*, 2008; Basu *et al.*, 2007; Hadley *et al.*, 2003; Newhouse and McClellan, 1998; among many others)

- **Exponential conditional mean models** (*e.g.* Manning *et al.*, 2005; Gilleskie and Mroz, 2004; among many others)

- **Generalized linear models** (*e.g.* Hill and Miller, 2010; Basu and Rathouz, 2005; Manning *et al.*, 2005; Buntin and Zaslavsky, 2004; Manning and Mullahy, 2001; among many others)

However, there are very few papers in Health Economics that take into account a spatial structure of discrete data. Among these, Lee and Cohen (1985) used an MNL model to describe the spatial distribution of the hospital utilization with the conclusion that the trip time was the most significant regressor for determining the hospital building choice. They used a *Nakanishi estimation method* compared with a *pseudo-Bayes approach* for dealing with zero observations. After about ten years, first attempts to consider a spatial autoregressive parameter in a discrete choice model have been seen. Bolduc *et al.* (1996b) estimated a Spatial Autoregressive Multinomial Probit (SAMNP) model of the choice of location by general practitioners for establishing their initial practice. The hybrid MNP model approximates the correlation among the utilities of the different locations using a first-order spatial autoregressive [SAR(1)] process based on a distance decaying relationship. They used a *maximum likelihood simulated (MSL) estimation* to describe the effect of various incentive measures introduced in Québec (Canada) to influence the geographical distribution of physicians across 18 regions. The price of medical services was treated as an exogenous variable, while the number of working hours and the quality and quantity of medical services were treated as endogenous ones. Data referred to 1976-1988 period with subperiods before and after the introduction of these measures. Results showed that incentive policies have had effects on physicians' location choices relating on the price and income effects. It was also provided a comparison between the standard MNP model and the Spatial MNP (SMNP) model, concluding in favor of the last one. In a related work, Bolduc *et al.* (1997) utilized the same Health data to compare the MSL-based estimation with choice probabilities simulated via GHK simulator and Gibbs sampling. As we said in paragraph 3, no significant difference in the estimates was found. An interesting work was that of Geweke *et al.* (2003), in which they analyzed hospital quality with an Endogenous Binary Probit (EBP) model characterized by nonrandom selection. They controlled for hospital selection using a model in which distances between the patient's residence and alternative hospitals were key exogenous variables. Mortality rates in patient discharge records were widely used. Data were collected from a sample of 74,848 Medicare patient admitted to 114 hospitals in Los Angeles County from 1989 through 1992 with a diagnosis of pneumonia. Results suggested that the smallest and largest hospitals exhibit higher quality than other hospitals. In a case-control study, De Iorio and Verzilli (2007) proposed a Bayesian multivariate probit model which flexibly accounts for the local spatial correlation between makers to identify patterns of genetic variation that differ across cases and controls. They used real data from the CYP2D6 region that has a confirmed role in drug metabolism. Reich and Bandyopadhyay (2010) developed a spatial multivariate model in a study on general periodontal Health with data collected during a periodontal exam. Studies based on *geo-additive probit models* (*see e.g.* Kammann and Wand, 2003) are for example those of Kandala *et al.* (2006) and Khatab and Fahrmeir (2009). The former examined the spatial distribution of observed diarrhea and fever prevalence in Malawi by using individual data for 10,185 children from the 2000 Malawi Demographic and Health Survey (DHS), the latter analyzed the impact of risk factors and the spatial effects on the latent variable "Health status" of a child less than 5 years of age by using the 2003 Egypt DHS. Leonard *et al.* (2009) used a spatial random effects probit model with which they analyzed the probability that a household from a region in Tanzania can recall another illness episode as a function of the characteristics of the illness, the location and type of Health care chosen and the outcome experienced. They collected data from an interview of 502 randomly selected households from 22 villages in 20 wards of the same region. Bukenya *et al.* (2003) used a Spatial Ordered Probit (SOP) model to examine the relationship existing between quality of life (QOL), Health and several socioeconomic variables, by utilizing a sample of more than 2000 residents in 21 counties of West Virginia and by generating spatial data geocoding survey respondents' addresses and hospital locations. They used a *maximum likelihood (ML) estimation*. Alexander et al. (2000) proposed a spatial model with negative binomial distribution to analyze a human parasite which causes a particular human disease, by utilizing individual count data from a province of Papua New Guinea and by adopting a *Bayesian approach with MCMC* (Metropolis-Hastings algorithm). The main objective was to identify the regions with augmented infection risk that can show environmental risk factors in a pretreatment situation. Best *et al.* (2000) used a Spatial Poisson model to analyze the traffic pollution effects on respiratory disorders in children by estimating parameters with *Bayesian approach and MCMC with data augmentation* (Gibbs sampling and Metropolis-Hastings algorithm). Data came from European Small-Area Variations in Air Quality and Health (SAVIAH) study and from annual average $NO_2$ concentrations. Congdon *et al.* (2007) proposed a Generalized Spatial Structural Equation Model (spatial SEM) for the impact on area Health referral counts of spatially correlated latent constructs. This model is generalized because it take into account both indicator-based and residual factors (Liu *et al.*, 2005), and it estimated by *Bayesian method with MCMC techniques*. As we said in paragraph 2.1 (subsection c), Bhat and Sener (2009) proposed a *copula-based approach* to estimate a spatial binary logit model. Their study was focused on teenagers' physical activity participation levels, a subject of considerable interest in the public Health as well as in other fields. Since physical activity is an inherent part of a Healthy lifestyle with the potential to increase the quality of life (QOL) and years of life, the model was used to examine the factors that influence whether or not a teenager participates in physical activity during the course of a day. The primary source of data is the 2000 San Francisco Bay Area Travel Survey (BATS) which collected detailed information on individual and household socio-demographic and employment-related characteristics from over 15000 households in the Bay Area. Finally, Neelon *et al.* (WP - 2011) proposed a spatial Poisson hurdle model for exploring geographic variation in emergency department (ED) visits while accounting for zero inflation. The model consists of a Bernoulli process which models the probability of any ED use and a truncated Poisson process which models the number of ED visits given use. The model has also a hierarchical structure that incorporates patient- and area-level covariates, as well as spatially correlated random effects for each areal unit. Spatial random effects are modeled in a Bayesian framework via a bivariate conditionally autoregressive (CAR) prior, which introduces dependence between the components and provides spatial smoothing and sharing of information across regions.

## 6. CONCLUSIONS

The main purpose of this paper consisted in reviewing the different methodological solutions in Spatial Discrete Choice models as they appeared in several applied fields by placing an emphasis on the Health Economics applications. We distinguished between different types of discrete choice models according as the dependent variable assumed two or more than two outcomes (*i.e.* binary vs. multinomial, ordered and count data), the estimation method (*i.e.* classical vs. Bayesian) and the applied filed (*i.e.* health vs. all the others). Papers that account for spatial autocorrelation in the discrete or limited dependent variable are still not many. One of the most important reasons for the relatively scarce diffusion of these models is their complexity, often requiring a multidimensional integration to estimate the set of parameters with a Full ML approach. As a consequence, an increasing attention has been placed on Bayesian inference methods (*i.e.* Gibbs sampling), as well as on semiparametric and nonparametric techniques (*i.e.* McDonald and McMillen, 2004), as computational solutions to estimate Spatial Discrete Choice models. Moreover, we found that only a small number of papers introduce the above-mentioned models in Health Economics, emerging the need to introduce the concept of spatial autocorrelation in this applied field.